\documentclass[11pt,twoside]{article}

\usepackage{asp2006}
\usepackage{graphicx}
\usepackage{epsf}
\usepackage{psfig}
\usepackage{lscape}
\begin{document}
\title{Amplitudes of High-Degree p-Modes in the Quiet and Active Sun}   %%%
%Fill in title

%\author{O. Burtseva\altaffilmark{1}, S. C. Tripathy\altaffilmark{1}, F.
%Hill\altaffilmark{1}, S. Kholikov\altaffilmark{1}, N.-E.
%Raouafi\altaffilmark{1}, and C. Lindsey\altaffilmark{2}}
%\altaffiltext{1}{National Solar Observatory, 950 N. Cherry Ave., Tucson, AZ
%85719, USA}
%\altaffiltext{2}{North West Research Associates, 3380 Mitchell Lane,
%Boulder, CO 80301, USA} %%% Fill in author

\author{O. Burtseva, S. C. Tripathy, F. Hill, S. Kholikov, N.-E. Raouafi}
\affil{\small{National Solar Observatory, 950 N. Cherry Ave., Tucson, AZ
85719, USA}}
%%% Fill in author affiliations
\author{C. Lindsey}   %%% Fill in author names
\affil{\small{North West Research Associates, 3380 Mitchell Lane, Boulder,
CO 80301, USA}} %%% Fill in author

\begin{abstract} We investigate mode amplitudes in the active and quiet Sun
in both maximum and minimum phases of the solar activity cycle. We confirm
previous studies showing that $p$-mode amplitudes at solar minimum are higher
than at solar maximum. We mask active regions of a certain magnetic field
strength and compare the masked and unmasked acoustic power. After applying
the masks, the preliminary analysis indicates that the amplitude decreases
over all degrees during solar minimum, compared to the unmasked case, while
at solar maximum the amplitude first decreases up to $\ell$ $\sim$ 300 and
then increases at higher degrees.  \end{abstract}

\section{Motivation}   

Both global and local analyses of intermediate- and high-degree modes
indicate that the mode amplitudes and the solar activity level are
anti-correlated \citep[e.g.][]{chaplin2000} and strongly depend on the local
magnetic flux \citep{rajaguru2001,howe2004}. This work addresses the
following questions: How different are the mode amplitudes in active and
quiet sun at minimum and maximum of the solar activity cycle? Does the quiet
sun have more acoustic power at solar minimum in comparison with solar
maximum or vice versa? These questions are addressed by masking active
regions above a certain magnetic field strength and by comparing the masked
and unmasked acoustic power at minimum and maximum phases of the solar
activity cycle.

\section{Data Analysis}   

We analyzed the following data sets: 

\begin{itemize}

\item[-] 10-hr time series of Michelson Doppler Imager (MDI) Dopplergrams, one
at solar minimum (May 24, 1996) and at solar maximum (September 9, 2000).  

\item[-] Eleven 10-hr long Global Oscillation Network Group (GONG) Dopplergram
time series around solar minimum (June 02-13, 2007) and solar maximum (December
03-14, 2001). 

\item[-] MDI full-disk magnetograms to identify active regions and create masks.
An example of the mask used in this work is shown in Figure 1. The mask consists
of zeros (black) in active regions and ones (white) otherwise. 

\end{itemize}

\begin{figure}
\centering
\includegraphics[width=0.95\linewidth]{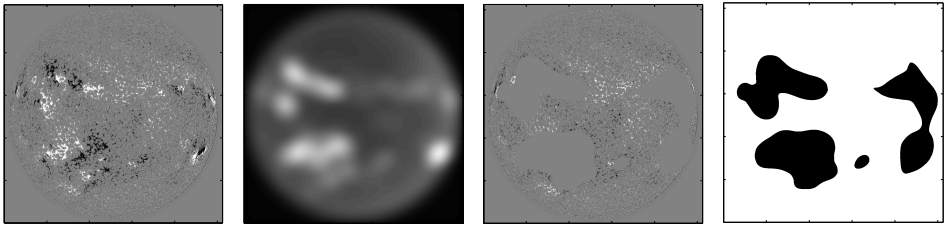}
\caption{\textit{From left to right}: (i) An MDI magnetogram from Sept. 9, 2000;
(ii) Smoothed unsigned magnetogram averaged over time; (iii) Masked magnetogram
obtained from the smoothed image with a cut-off value of 20 G showing only the
quiet regions; (iv) Mask of strong field regions.}
\end{figure}

We combine two analysis methods (i) the standard ring-diagram technique from
the GONG pipeline \citep{corbard2003} where patches within 30$\deg$ of disk
center are analyzed, and (ii) the asymmetric-peak-profile fitting of the power
spectrum ($\ell$-$\nu$ diagram) \citep{nigam1998}. In this procedure,
Dopplergrams are remapped, tracked, filtered with a 15 min running mean, and
decomposed into Spherical Harmonic coefficients to construct an $\ell$-$\nu$
diagram. In this study, we restrict the analysis to \textit{p}1-\textit{p}4
ridges as a function of the mode degree $\ell$. 

\begin{figure}[!hb] \centering
\includegraphics[width=0.70\linewidth]{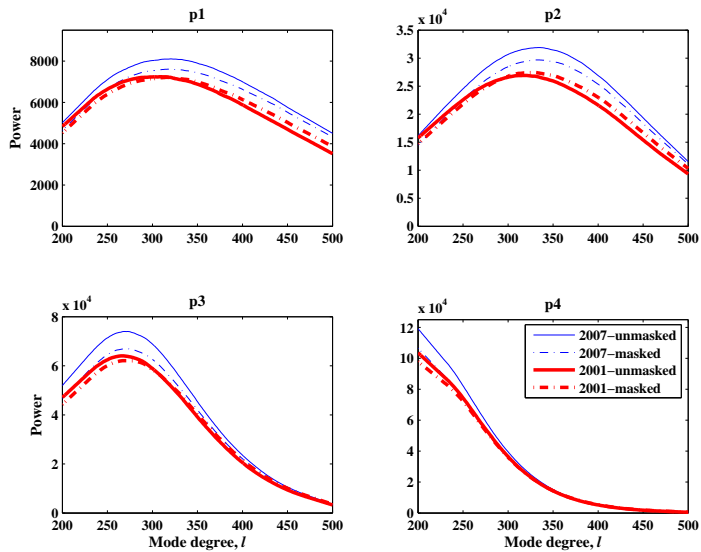}     
\caption{Amplitudes of \textit{p}1-\textit{p}4 modes as a function of the mode
degree obtained from power spectrum analysis of 10-hr time series of GONG
Dopplergrams, averaged over 11 days each in 2001 (solar maximum) and 2007
(solar minimum). The line styles are defined in the right bottom panel.}
\end{figure}

\begin{figure}
\centering
\includegraphics[width=0.82\linewidth]{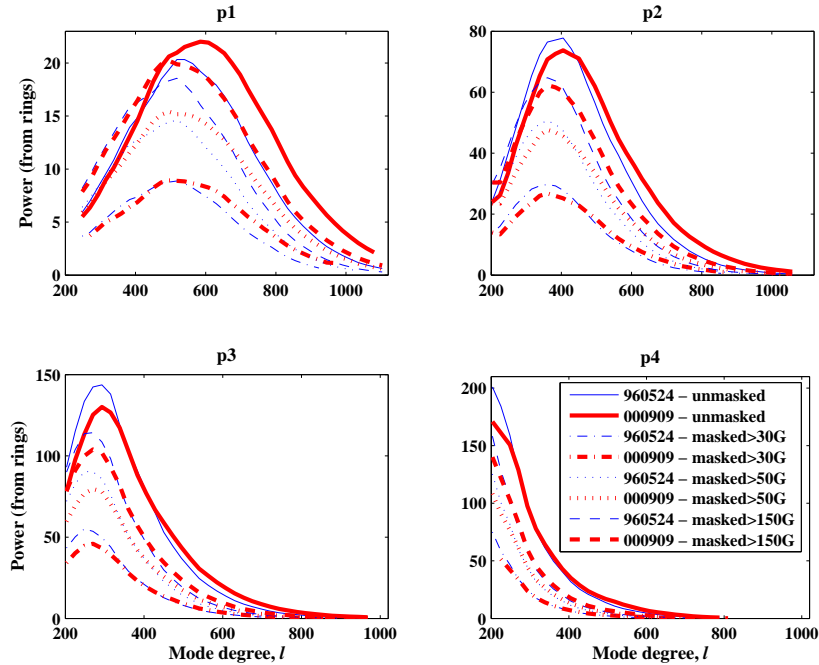}
\caption{Amplitudes of \textit{p}1-\textit{p}4 modes as a function of the mode
degree obtained from 10 hr of MDI data by masking the pixels with unsigned
magnetic field above certain threshold values. The line styles are defined in
the right bottom panel.}
\end{figure}

\section{Results and Discussion}   

From power-spectrum and ring analysis applied to 10-hr MDI data sets, we
observe that the amplitude of \textit{p}1-\textit{p}4 modes at solar minimum is
lower than that at solar maximum up to $\ell$ $\sim$ 300 - 400, and higher for
$\ell >$ 400. The same trend is noticed when we apply the mask but the
amplitudes decrease for all modes in comparison with the unmasked case. The
analysis showed significant day-to-day variations. In order to minimize these
systematic variations, we averaged the data over a period of 11 days each in
2001 and 2007. 

In Figure 2, the averaged amplitudes of \textit{p}1-\textit{p}4 modes as a
function of $\ell$ obtained from GONG in 2001 and 2007 are presented. The
results confirm that the amplitude at solar minimum is higher than at solar
maximum for the entire analyzed $\ell$-range. After masking the active regions,
we observe that the amplitude at solar minimum decreases over all degrees,
while at solar maximum the amplitude decreases up to $\ell$ $\sim$ 300 and
increases at higher degrees.

In an attempt to understand the cause of the decrease in amplitude when the
mask is applied, we excluded active regions by masking pixels with unsigned
magnetic field above a certain threshold levels and reprocessed the
Dopplergrams through the ring-diagram technique. The results in Figure 3 show
that the mode amplitudes decrease with removal of flux; the mode amplitudes
became smaller as more activity was removed. These results are consistent with
those obtained from the masking of the active region and discussed above. 

\begin{figure}
\centering
\includegraphics[width=0.78\linewidth]{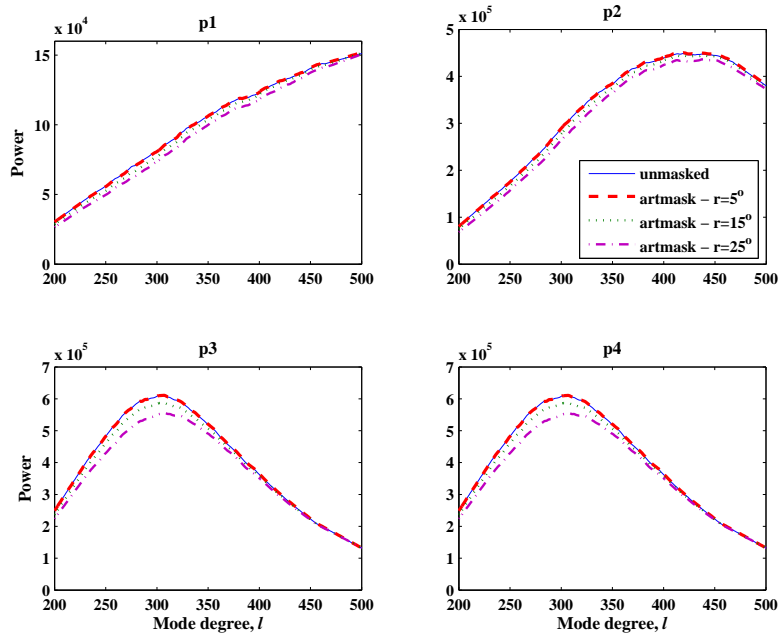}
\caption{Amplitudes of \textit{p}1-\textit{p}4 modes as a function of the mode
degree obtained from the power spectrum analysis of MDI data (Nov. 12, 2006).
Here we construct circular artificial masks of different radius. The line
styles are defined in the right top panel.}
\end{figure}

In order to understand the changes in amplitude, we performed a numerical
simulation where we applied a circular artificial mask of three different radii
to one day of MDI data (November 12, 2006). The results obtained from fitting
of the power spectrum are shown in Figure 4. It appears that the variation in
the amplitude could be due to the effect of the mask alone. We plan to extend
this study with artificial masks of different sizes and locations on the solar
disk. We further plan to carry out center-to-limb corrections of the
line-of-sight velocity to understand its effect on the mode amplitudes. 

\acknowledgements 
The National Solar Observatory is operated by AURA, Inc. under a cooperative
agreement with the National Science Foundation. SOHO is a mission of
international cooperation between ESA and NASA.

\end{document}